\documentclass[twocolumn,10pt]{IEEEtran}
\IEEEoverridecommandlockouts
\usepackage{amsmath}
\usepackage{amsfonts}
\usepackage{amssymb}
\usepackage{amsthm}
\usepackage{graphicx}
\usepackage{psfrag}
\usepackage{cite}
\usepackage{algorithm}
\usepackage{algorithmic}
\usepackage{color}
\usepackage{float}%
\usepackage{epsfig,array,multicol,verbatim}
\usepackage{subfigure}
\usepackage{cases}
\usepackage{setspace}

\theoremstyle{plain} \newtheorem{theorem}{Theorem}
\theoremstyle{plain} 
\theoremstyle{plain} 
\theoremstyle{remark} \newtheorem{remark}{Remark}
\theoremstyle{remark} \newtheorem{lemma}{Lemma}
\theoremstyle{plain} \newtheorem{definition}{Definition}

\begin{document}
\title{Multi-Round Contention in Wireless LANs with Multipacket Reception}
\author{Ying~Jun~(Angela)~Zhang,~\IEEEmembership{Member,~IEEE}\\
Dept. of Information Engineering, The Chinese University of Hong
Kong\\Email: yjzhang@ie.cuhk.edu.hk
\thanks{This work was supported in part by the Competitive Earmarked
Research Grant (Project Number 418707) established under the
University Grant Committee of Hong Kong, and the Direct Grant for Research (Project Number 2050370) established by The Chinese University of Hong Kong.}}

\maketitle

\begin{abstract}
Multi-packet reception (MPR) has been recognized as a powerful capacity-enhancement technique for random-access wireless local area networks (WLANs). As is common with all random access protocols, the wireless channel is often under-utilized in MPR WLANs. In this paper, we propose a novel multi-round contention random-access protocol to address this problem. This work complements the existing random-access methods that are based on single-round contention. In the proposed scheme, stations are given multiple chances to contend for the channel until there are a sufficient number of ``winning" stations that can share the MPR channel for data packet transmission. The key issue here is the identification of the optimal time to stop the contention process and start data transmission. The solution corresponds to finding a desired tradeoff between channel utilization and contention overhead. In this paper, we conduct a rigorous analysis to characterize the optimal strategy using the theory of optimal stopping. An interesting result is that the optimal stopping strategy is a simple threshold-based rule, which stops the contention process as soon as the total number of winning stations exceeds a certain threshold. Compared with the conventional single-round contention protocol, the multi-round contention scheme significantly enhances channel utilization when the MPR capability of the channel is small to medium. Meanwhile, the scheme automatically falls back to single-round contention when the MPR capability is very large, in which case the throughput penalty due to random access is already small even with single-round contention.
\end{abstract}

\section{Introduction}
\subsection{Motivation and Contributions}
In  random-access wireless networks, such as IEEE 802.11 wireless local area networks (WLAN), stations share a common medium through contention-based medium access control (MAC). Most of what we know about WLAN is based on the conventional collision model, where packet collisions occur when two or more stations transmit at the same time \cite{Bianchi:00, Song:03}.
With advanced PHY-layer signal processing techniques, it is possible for an access point (AP) to detect multiple concurrently transmitted packets through, for example, multiuser detection (MUD) techniques \cite{Verdu:98, Tong:01}. This new collision model, referred to as multi-packet reception (MPR), opens up new possibilities for drastically enhancing the capacity of WLANs. Our prior work in \cite{Zhang:09} shows that the throughput of WLANs scales super-linearly with the MPR capability of the channel.

With MPR, up to $M$ stations can transmit at the same time without causing collisions, where $M$ is referred to as the MPR capability of the channel. An immediate question is how the MAC should be redesigned to fully utilize the advantages of MPR. In \cite{Zheng:06, Zhang:09}, we derived the optimal transmission probability and backoff exponent that maximize the throughput of MPR WLAN. With the optimal transmission probability, system throughput is greatly enhanced compared with that in traditional single-packet reception (SPR) WLANs.

One observation from our prior work, however, is that the MPR channel is still under-utilized from time to time even when the optimal transmission probability is adopted. In other words, the channel is not always fully occupied by $M$ concurrent packet transmissions during the data transmission phase. This is because the current  MAC protocols are based on a ``single-round contention" framework. There is essentially only \emph{one} contention round for each data transmission phase. For example, in the DCF RTS/CTS access mode, a data transmission phase follows immediately as long as there is \emph{one} successful RTS contention. Similarly, in the DCF basic access mode, data packet transmission also serves the purpose of channel contention, which implies that there is only one round of contention for each data transmission. Due to the random-access nature, the number of stations contending for the channel at a time is a random variable. Hence, the channel is unavoidably under-utilized when there are less than $M$ stations contend for the channel simultaneously. As such, enhancing the capacity of MPR WLANs beyond what is currently achievable remains a challenging problem.

This paper proposes a novel \emph{multi-round contention} random-access protocol to address the problem. With the multi-round contention framework, more contention rounds are executed before data transmission if the number of stations that have already won the channel contention is small. Intuitively, the more contention rounds, the more likely that the channel is fully packed with $M$ concurrent packet transmissions in the data transmission phase. On the other hand, more contention rounds leads to higher channel-contention overhead. Finding the desired tradeoff between channel utilization and contention overhead boils down to deciding when to terminate the contention rounds and start data transmission. In this paper, we conduct a rigorous analysis to characterize the optimal strategy using the theory of optimal stopping \cite{Ferguson}. The key contributions of this paper are summarized in the following.

\begin{itemize}
  \item We show that the problem of finding the optimal stopping strategy that maximizes the system throughput is equivalent to the problem of maximizing the rate of return (MR), a subclass of optimal stopping problems.
  \item By exploiting the monotone nature of the problem, we prove that the optimal stopping strategy for multi-round contention is a simple threshold-based rule. Specifically, it is optimal to terminate the contention rounds as soon as the total number of stations that have succeeded in channel contention exceeds a certain threshold, regardless of the number of contention rounds that have already been executed.
  \item Based on the analysis, the maximum throughput that is achievable in MPR networks with multi-round contention is derived. In particular, network throughput is maximized when the stopping threshold and the transmission probability of stations are jointly optimized. Our results show that multi-round contention drastically enhances the channel utilization compared with networks with single-round contention, especially for small to moderate $M$, which is the case in most practical situations. This analysis complements our work in \cite{Zhang:09} that has focused on MPR WLANs with single-round contention.
 \item For practical implementation, we propose a multi-round contention protocol which only requires minor revisions to the current IEEE 802.11 DCF.
\end{itemize}

\subsection{Related Work}

In related work, \cite{Chan:04} attempts to enhance the utilization of MPR channels by allowing stations to count down and transmit as long as there are less than $M$ ongoing transmissions in the air. To do this, one key assumption is that a station is able to detect the number of ongoing transmissions using an energy detector. This assumption, however, is not valid in wireless networks, where the received energy from each transmitting station is random and unknown a priori.

The theory of optimal stopping has been widely studied in the fields of statistics, economics, and mathematical finance since 1960's \cite{Chow:71}. It was not until very recently that optimal stopping theory started to find application in wireless networks. In \cite{Jia:08}, the tradeoff between the spectrum access opportunity and spectrum sensing overhead in cognitive radio systems is formulated as a finite-horizon optimal stopping problem, which is solved using backward induction. Likewise, a finite-horizon optimal stopping problem is formulated in \cite{Ai:08} to derive an optimal next-hop selection strategy in multi-hop ad hoc networks. The problem of maximizing the rate of return (MR) was applied to opportunistic scheduling in ad-hoc networks in \cite{Zheng:09} and opportunistic spectrum access of cognitive radio networks in \cite{Huang:08}. Notably, the application of optimal stopping theory in wireless systems is still at its infancy stage. Our work in this paper is an attempt to introduce it to wireless random-access networks.

The rest of this paper is organized as follows. The system model of MPR WLAN with multi-round contention is introduced in Section II, where we also show that the problem of maximizing network throughput can be formulated as the problem of MR, a subclass of infinite-horizon optimal stopping problems. A multi-round contention protocol as a minor amendment of IEEE 802.11 RTS/CTS access mode is presented in the same section. A preliminary on optimal stopping theory is presented in Section III. In Section IV, we prove that the optimal stopping strategy for multi-round contention in MPR WLAN is a simple threshold-based strategy. The network throughput as well as the lower bound on the maximum throughput is derived in Section V. In Section VI, we investigate the throughput improvement of multi-round contention MAC over existing single-round contention MAC through numerical results. Finally, the paper is concluded in Section VII. For the convenience of the readers, the notations in this paper are summarized in Table \ref{tab:notation}.

\begin{table}
\caption{Summary of Notations}\label{tab:notation}
\centering
\small \begin{tabular}{|c|c|}
  \hline
  $K$ & Number of mobile stations \\
  \hline
  $M$ & MPR capability of the channel  \\
  \hline
  $X_i$ & Number of winning stations per contention round\\
  \hline
  $\lambda$ & Average number of transmission attempts per generic time slot\\
  \hline
  $N$ & Stopping time \\
  \hline
  $T_N$ & Duration of a super round \\
  \hline
  $Y_N$ & Data payload transmitted in a super round \\
  \hline
  $N^*$ & The optimal stopping time \\
  \hline
  $N_1$ & The stopping time by the one-stage look-ahead rule \\
  \hline
  $\theta$ & The threshold in the one-stage look-ahead rule \\
  \hline
  $S$ & System throughput \\
  \hline
  $B_L$ & Throughput lower bound \\
  \hline
  $S_c$ & Throughput for multi-round contention with carry-over \\
  \hline
\end{tabular}
\end{table}

\section{System Model and Problem Formulation}
\subsection{Multi-round Contention and Problem of Maximizing Rate of Return}
\begin{figure}[h!]
\centering
\includegraphics[width=0.5\textwidth]{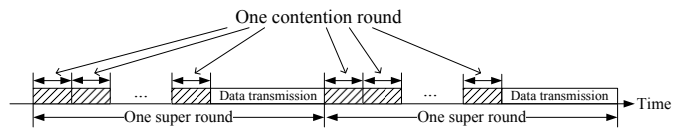}
\caption{Multi-round contention} \label{fig:multiround}
\end{figure}

We consider a fully connected network with $K$ mobile stations transmitting to an AP. The transmission of stations is coordinated by a random-access protocol. We assume that the AP has the capability to decode up to $M$ simultaneous packet transmissions, be it contention packets or data packets. Interested readers are referred to Section V in \cite{Zhang:09} for a practical protocol to implement MPR in random access networks.  A sketch of the multi-round contention mechanism is illustrated in Fig. \ref{fig:multiround}. The precise model will be made concrete in the next subsection, where we propose a multi-round contention protocol as a minor amendment of IEEE 802.11 RTS/CTS mechanism.

\begin{figure*}[t!]
\centering
\includegraphics[width=0.8\textwidth]{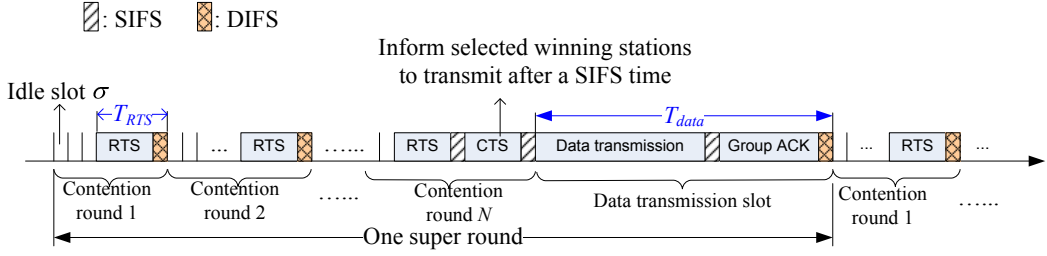}
\caption{Multi-round contention for IEEE 802.11 WLAN} \label{fig:802.11multiround}
\end{figure*}

In Fig. \ref{fig:multiround}, the time axis is divided into contention rounds and data transmission slots. The period between the ends of two neighboring data transmission slots is referred to as a super round.  Stations transmit a small contention packet with probability $\tau$ in each contention round. If there are no more than $M$ stations contending for the channel in the same contention round, then all contention packets are decoded and the stations are said to have won the contention. Otherwise, if more than $M$ stations send contention packets in the same contention round, a collision occurs and none of them win the channel contention.

Let $\{X_1, X_2, \cdots\}$ denote a sequence of random variables representing the number of winning stations in each contention round. Obviously, $0\leq X_i\leq M$. For each $i=1, 2, \cdots$, after observing $X_1=x_1, X_2=x_2, \cdots, X_n=x_n$, we may stop the contention and transmit $\sum_{i=1}^n x_i$ data packets in the data transmission slot if $\sum_{i=1}^n x_i\leq M$; only $M$ winning stations will be selected to transmit if $\sum_{i=1}^n x_i>M$. In other words, the number of transmitting stations is
\begin{equation}
y_n(x_1, \cdots, x_n)=\max(\sum_{i=1}^n x_i, M).
\end{equation}
Instead of stopping the contention at the $n^{th}$ round, we may also continue and observe $X_{n+1}$, hoping that $y_{n+1}(x_1, \cdots, x_{n+1})$ will be much larger than $y_n$. Of course, this is at the risk of wasting more time on contention without getting a reasonably larger $y_{n+1}$ in return.

A stopping rule $\phi$  determines the stopping time $N$ based on the sequence of observations $\mathbf{X}=(X_1, X_2, \cdots)$. Note that $N$ is random, as it is a function of random variables $\mathbf{X}$. Different realizations of observations may lead to different stopping time. The system throughput can then be calculated as
\begin{eqnarray}\label{eqn:rateofreturn}
S_\phi&=&\frac{\mathrm{E}_\mathbf{X}[\text{Data payload transmitted in one super round}]}{\mathrm{E}_{\mathbf{X}}[\text{Duration of a super round}]}\nonumber\\
&=&\frac{\mathrm{E}_\mathbf{X}[Y_N]}{\mathrm{E}_\mathbf{X}[T_N]},
\end{eqnarray}
where $Y_1, Y_2, \cdots$ is a sequence of random variables with realizations being $y_1, y_2, \cdots$. $T_N$
is the random variable representing the total amount of time spent to obtain a return of $Y_N$.

Let $\mathcal{C}$ denote the class of stopping rules with
\begin{equation}
\mathcal{C}=\{N:N\geq1, \mathrm{E}[T_N]<\infty\}.
\end{equation}
Our purpose is to find the optimal stopping rule $N^*\in\mathcal{C}$ that maximizes the system throughput $\frac{\mathrm{E}_\mathbf{X}[Y_N]}{\mathrm{E}_\mathbf{X}[T_N]}$. In optimal stopping theory, this problem is referred to as the problem of MR.

\subsection{Multi-round Contention in IEEE 802.11 WLAN}
Having introduced the general framework of multi-round contention, we now propose a multi-round contention protocol based on IEEE 802.11 RTS/CTS access mode. Note that the problem formulated in the preceding subsection and the analysis in later sections are general and not restricted to the protocol proposed in this subsection.

In IEEE 802.11, the transmission of stations is coordinated by an exponential backoff (EB) mechanism. The EB mechanism adaptively tunes the transmission probability of a station according to the traffic intensity of the network. It works as follows. At each packet transmission, a station sets its backoff timer by randomly choosing an integer within the range $[0, W-1]$, where $W$ is the size of the contention window. The backoff timer freezes when the channel is busy and is decreased by one following each time slot when the channel is idle. The station transmits a packet from its buffer once the backoff timer reaches zero. At the first transmission attempt of a packet, $W$ is set to $W_0$, the minimum contention window. Each time the transmission is unsuccessful, $W$ is multiplied by a backoff factor $r$. That is, the contention window size $W_j = r^jW_0$ after $j$ successive transmission failures.

The multi-round contention protocol is illustrated in Fig. \ref{fig:802.11multiround}. A station transmits an RTS packet when its backoff timer reaches zero. Previous work in \cite{Bianchi:00, Song:03, Zhang:09} has shown that the backoff process yields an equivalent transmission probability $\tau$ at which a station transmits in a generic (randomly chosen) time slot. When the number of stations, $K$, is large, it is reasonable to assume that the number of transmissions in a generic time slot follows a Poisson distribution with parameter $\lambda=K\tau$ \cite{Zhang:09}. That is,
\begin{equation}\label{eqn:attempt}
\Pr\{k~\text{stations transmit in a generic time slot}\}=\frac{\lambda^k}{k!}e^{-\lambda}.
\end{equation}
 If no more than $M$ stations transmit at the same time, then the contention is successful and these stations are marked as winning stations. Otherwise, a collision occurs and there are zero winning stations. From \eqref{eqn:attempt}, it can be shown that the number of winning stations $X_i$ follows the following distribution:
\begin{equation}\label{eqn:X-distribution}
\Pr\{X_i=k\}=
\begin{cases}
\frac{\lambda^k}{k!(e^\lambda-1)}&1\leq k\leq M\\
\sum_{j=M+1}^\infty \frac{\lambda^j}{j!(e^\lambda-1)}& k=0\\
0&\text{otherwise}
\end{cases},
\end{equation}
with the expectation being
\begin{equation}\label{eqn:meanX}
\mathrm{E}[X]=\frac{\lambda}{1-e^{-\lambda}}\sum_{k=0}^{M-1}\frac{\lambda^k x^{-\lambda}}{k!}.
\end{equation}

After observing the outcome of the contention, the AP determines whether to stop the contention rounds according to the optimal stopping strategy. It keeps silent if it decides not to stop the contention rounds. According to IEEE 802.11 DCF, other stations will then continue to count down after sensing the channel idle for a DIFS (DCF interframe space) time and contend for the channel when their counter values reach zero. If the AP decides to stop the contention rounds, it will randomly select, from all winning stations, at most $M$ stations for data packet transmission. Note that if the total number of winning stations in this super round does not exceed $M$, then all of them will be select. This decision is broadcasted to all mobile stations through a CTS packet after a SIFS interval. Then, the selected winning stations send their data packets. After that, the AP responds with a group ACK, indicating which data packets have been received successfully.

The stations that have contended but are not notified to transmit data by the CTS packet regard themselves as having encountered a collision, and consequently multiply their contention window by $r$ and back off. Note that the collision can be either an actual one that occurs when more than $M$ stations transmit together in a contention round, or a virtual one that occurs to winning stations that are not selected by the AP when the total number of winning stations exceeds $M$  by the end of the last contention round. This protocol falls back to the traditional single-round contention protocol if the AP always terminates contention after the first successful contention round.

Under the analytical framework described in the last subsection, we regard one RTS contention \emph{including} the preceding idle slots and the succeeding interframe spaces as one contention round, as illustrated in Fig. \ref{fig:802.11multiround}. Likewise, the ``data transmission slot'' contains the data packet transmission, the group ACK, together with the interframe spaces. Let $T_{RTS}$ and $T_{data}$ be the durations defined in Fig. \ref{fig:802.11multiround}, i.e.,

\begin{equation}\label{eqn:RTSCTS}
T_{RTS}=RTS+DIFS,
\end{equation}
\begin{equation}\label{eqn:Tdata}
T_{data}=T_H+\frac{L}{R}+T_{ACK}+SIFS+DIFS,
\end{equation}
where $T_H$ denotes the transmission time of a packet header, $L$ denotes the payload length of a packet, $R$ denotes the data transmission rate, and $T_{ACK}$ denotes the time duration of a group ACK packet. The acronyms (i.e., RTS, CTS, SIFS, DIFS, ACK) represent the corresponding time duration specified in the IEEE 802.11 standard.

If the contention phase stops at the $N^{th}$ round, then
\begin{equation}\label{eqn:TN}
T_N=NT_{RTS}+\sum_{i=1}^N I_i\sigma+CTS+2SIFS-DIFS+T_{data},
\end{equation}
where $\sigma$ is the length of a idle slot and $I_i$ is the number of idle slots preceding the  RTS packet in the $i^{th}$ contention round. From \eqref{eqn:attempt}, it can be seen that a slot is idle with probability $e^{-\lambda}$ when $K$ is large. Therefore, $I_i$ follows a geometric distribution with mean value
\begin{equation}\label{eqn:meanI}
m_I=\mathrm{E}[I]=\frac{e^{-\lambda}}{1-e^{-\lambda}}.
\end{equation}
The term $CTS+2SIFS-DIFS$ in \eqref{eqn:TN} is due to the fact that the duration of the last contention round is statistically different from others.

System throughput defined in \eqref{eqn:rateofreturn} can now be written as
\begin{eqnarray}\label{eqn:thrput802.11}
S_\phi&=&\frac{\mathrm{E}_\mathbf{X}[Y_N]}{\mathrm{E}_{\mathbf{X},I}[T_N]}\nonumber\\
&=&\frac{\mathrm{E}_\mathbf{X}[\min(\sum_{i=1}^N X_i,M)]}{\mathrm{E}_{\mathbf{X},I}[NT_{RTS}+\sum_{i=1}^N I_i\sigma+B]}\nonumber\\
&=&\frac{\mathrm{E}_\mathbf{X}[\min(\sum_{i=1}^N X_i,M)]}{\mathrm{E}_{\mathbf{X}}[N(T_{RTS}+m_I\sigma)+B]} \text{ packets/second}
\end{eqnarray}
where $B=CTS+2SIFS-DIFS+T_{data}$ is a constant invariant of the stopping criterion.

Intuitively, the decision whether to stop contention at a certain round could be based on the number of contention rounds that have already taken place, the number of stations that have already won the contention, or a combination of both. However, our analysis in Section IV reveals a somewhat surprising result:  The optimal stopping rule is solely based on the number of winning stations, regardless of how many contention rounds that have already been executed.

\section{Preliminary on Optimal Stopping Theory}
Before deriving the optimal stopping rule in the next section, we introduce in this section some definitions and theorems that will be useful in our later discussions. Theorem \ref{thm:MR} states that the problem of MR is equivalent to a stopping rule problem that aims to maximize the return $Y_N-\mu T_N$ for some $\mu$, where $Y_N$ and $T_N$ are as defined in \eqref{eqn:rateofreturn}.

\begin{theorem}[\cite{Ferguson}]\label{thm:MR}
(a) If $\sup_{N\in\mathcal{C}}\frac{\mathrm{E}_\mathbf{X}[Y_N]}{\mathrm{E}_\mathbf{X}[T_N]}=\mu$ and if the supremum is attained at $N^*$, then $\sup_{N\in\mathcal{C}}(\mathrm{E}_\mathbf{X}[Y_N]-\mu\mathrm{E}_\mathbf{X}[T_N])=0$ and the supremum is also attained at $N^*$.

(b) Conversely, if  $\sup_{N\in\mathcal{C}}(\mathrm{E}_\mathbf{X}[Y_N]-\mu\mathrm{E}_\mathbf{X}[T_N])=0$ for some $\mu$, then $\sup_{N\in\mathcal{C}}\frac{\mathrm{E}_\mathbf{X}[Y_N]}{\mathrm{E}_\mathbf{X}[T_N]}=\mu$. Moreover, if $\sup_{N\in\mathcal{C}}(\mathrm{E}_\mathbf{X}[Y_N]-\mu\mathrm{E}_\mathbf{X}[T_N])=0$ is attained at $N^*\in\mathcal{C}$, then $N^*$ is optimal for maximizing $\sup_{N\in\mathcal{C}}\frac{\mathrm{E}_\mathbf{X}[Y_N]}{\mathrm{E}_\mathbf{X}[T_N]}$.
\end{theorem}

\begin{remark}
$\mu$ is in fact the optimal rate of return which is equal to $\sup_{N\in\mathcal{C}}\frac{\mathrm{E}_\mathbf{X}[Y_N]}{\mathrm{E}_\mathbf{X}[T_N]}$.
\end{remark}

Theorem \ref{thm:MR} implies that to maximize the system throughput, we can alternatively solve a regular stopping rule problem that maximizes $Z_N$, where $Z_N=Y_N-\mu T_N$. It can be shown that the optimal stopping rule is the one that satisfies the following equation\footnote{The statement is valid when the optimal stopping rule exists \cite{Ferguson}, which can be proved for our particular problem. We omit the proof here for brevity.}.
\begin{eqnarray}\label{eqn:stoppingrule}
N^*&=&\min\big\{n\geq 1: \\
&& Z_n\geq \sup_{m\geq n}\mathrm{E}_\mathbf{X}[Z_m|X_1=x_1, \cdots, X_n=x_n]\big\}.\nonumber
\end{eqnarray}
 In other words, it is optimal to stop at a stage if the return at this stage is no less than the expected return of stopping at a future stage.


\begin{definition}[One-stage look-ahead rule]\label{def:1sla}
The one-stage look-ahead (1-sla) rule is the one that stops if the return for stopping at the stage is at least as large as the expected return of continuing one stage and then stop. Mathematically, the 1-sla rule is described by the stopping time
\begin{eqnarray}\label{eqn:1sla}
N_1 &= &\min\big\{n\geq 1:\\
&&Z_n\geq \mathrm{E}_\mathbf{X_{n+1}}[Z_{n+1}|X_1=x_1, \cdots, X_n=x_n]\big\}.\nonumber
\end{eqnarray}
\end{definition}

\begin{remark}
The 1-sla rule is not optimal in general. However, Definition \ref{dfn:monotone} and Theorem \ref{thm:1sla} show that $N^*=N_1$ when some conditions are satisfied.
\end{remark}

\begin{definition}\label{dfn:monotone}
Let $A_n$ denote the event $\{Z_n\geq\mathrm{E}[Z_{n+1}|X_1=x_1,\cdots, X_n=x_n]\}$. We say the stopping rule problem is monotone if $A_0\subset A_1\subset A_2\subset \cdots$. In other words, the problem is monotone if the one-stage look-ahead calls for stopping at stage $n$, then it will also call for stopping at all future stages no matter what the future observations turn out to be.
\end{definition}
\begin{theorem}\label{thm:1sla}
If $\lim_{n\rightarrow\infty}Z_n=Z_\infty$, $\mathrm{E}[\sup_n|Z_n|]<\infty$, and the problem is monotone, then the 1-sla rule is optimal.
\end{theorem}
\noindent\emph{Proof}: See Chapter 5.2 and 5.3 of \cite{Ferguson}.$\hfill\blacksquare$

\section{Optimal Stopping Rule for MPR WLAN with Multi-round Contention}

In this section, we analyze the optimal stopping rule that maximizes the system throughput of MPR WLAN with multi-round contention. In what follows, Lemma \ref{lem:threshold} shows that the 1-sla rule is a threshold based rule and the stopping time is solely determined by the number of stations that have already won the contention. Furthermore, it is proved in Lemma \ref{lem:1slaoptimal} that the 1-sla rule is the optimal stopping rule for our particular problem of maximizing system throughput of MPR WLAN.

\begin{lemma}\label{lem:threshold}
The 1-sla rule that maximizes $\frac{\mathrm{E}_\mathbf{X}[Y_N]}{\mathrm{E}_\mathbf{X}[T_N]}$ or, equivalently, $Y_N-\mu T_N$ is a threshold based rule that stops at the $n^{th}$ contention round as soon as $\sum_{i=1}^n X_i \geq\theta$. When the number of stations $K$ is large, $\theta$ is a fixed constant invariant with $n$.
\end{lemma}
\noindent\emph{Proof}:
Note that
\begin{eqnarray}
Z_N&=&Y_N-\mu T_N\\
&=&\min\big(\sum_{i=1}^NX_i, M\big)-\mu NT_{RTS}-\mu \sum_{i=1}^NI_i\sigma-\mu B\nonumber\\
&=&M-\big(M-\sum_{i=1}^NX_i\big)^+-\mu NT_{RTS}-\mu \sum_{i=1}^NI_i\sigma-\mu B,\nonumber
\end{eqnarray}
where $(\cdot)^+$ is equal to the argument if the argument is positive, and zero otherwise.

\begin{table*}[t!]
\begin{eqnarray}
N_1&=&\min\bigg\{n\geq 1: \big(M-\sum_{i=1}^n X_i\big)^+
-\mathrm{E}_{X_{n+1}}\bigg[\big(M-\sum_{i=1}^n X_i-X_{n+1}\big)^+\bigg|X_1=x_1, \cdots, X_n=x_n\bigg]
\leq \mu(T_{RTS}+m_I\sigma)\bigg\}\nonumber\\
&=&\min\bigg\{n\geq 1: \big(M-\sum_{i=1}^n X_i\big)-\mathrm{E}_{X_{n+1}}\bigg[\big(M-\sum_{i=1}^n X_i-X_{n+1}\big)^+\bigg|X_1=x_1, \cdots, X_n=x_n\bigg]
\leq \mu(T_{RTS}+m_I\sigma)\bigg\}\nonumber\\
&(a)\atop =&\min\big\{n\geq 1: M-\sum_{i=1}^nX_i\leq v_n\big\}\nonumber\\
&=&\min\big\{n\geq 1: \sum_{i=1}^nX_i\geq \theta_n\big\}\label{eqn:1slathreshold}
\end{eqnarray}
\end{table*}

The 1-sla rule described in \eqref{eqn:1sla} can now be rewritten as \eqref{eqn:1slathreshold} shown at the top of the next page, where $\theta_n=M-v_n$ and
\begin{equation}\label{eqn:vn}
v_n=\max\big\{u: u-\mathrm{E}_{X_{n+1}}[(u-X_{n+1})^+]\leq\mu(T_{RTS}+m_I\sigma)\big\}.
\end{equation}
When $K$ is large, the distribution of $X_i$ is identical for all $i$ (see \eqref{eqn:X-distribution}). In this case, $v_n$ and $\theta_n$ are invariant with $n$. If no confusion arises, the subscript $n$ will be omitted hereafter. It is obvious from \eqref{eqn:1slathreshold} that the 1-sla rule is a threshold based rule with a constant threshold $\theta$. $\hfill\blacksquare$
\begin{remark}
Equation (a) in \eqref{eqn:1slathreshold} are due to the fact that
$u-\mathrm{E}_{X_{n+1}}[(u-X_{n+1})^+]$
is an increasing function of $u$.
\end{remark}

\begin{lemma}\label{lem:1slaoptimal}
For MPR WLANs with multi-round contention, the stopping rule $N_1$ obtained by \eqref{eqn:1slathreshold} is the optimal solution to Problem \eqref{eqn:stoppingrule}. That is, $N_1=N^*$.
\end{lemma}
\noindent\emph{Proof}:
To prove Lemma \ref{lem:1slaoptimal}, we note that the return function $Z_n=Y_n-\mu n(T_{RTS}+m_I\sigma)-\mu B$ has the following properties in our particular problem.
\begin{equation}
\lim_{n\rightarrow\infty}Z_n=Z_\infty=-\infty,
\end{equation}
and
\begin{equation}
\mathrm{E}[\sup_n Z_n]\leq M-\mu(T_{RTS}+m_I\sigma)-\mu B<\infty.
\end{equation}
Furthermore, it can be seen from \eqref{eqn:1slathreshold} that the problem is monotone, because as long as the threshold $\theta$ is exceeded at a certain stage $n$ and the 1-sla rule calls for stopping, the threshold will always be exceeded at all future stages regardless of the future observations of $X$. Therefore, the 1-sla rule is the optimal stopping rule according to Theorem \ref{thm:1sla}. $\hfill\blacksquare$

\begin{theorem}\label{thm:threshold}
The optimal stopping rule that maximizes the throughput of MPR WLAN is a threshold based rule described as follows.
\begin{equation}\label{eqn:optimalstopping}
N^*=\min\big\{n\geq 1: \sum_{i=1}^nX_i\geq \theta\big\}.
\end{equation}
\end{theorem}
\noindent\emph{Proof}: Straightforward from Lemma \ref{lem:threshold} and Lemma \ref{lem:1slaoptimal}. $\hfill\blacksquare$

\begin{remark}
With MPR capability $M$, $Y_N$ is always upper bounded by $M$. In this case, increasing $\theta$ beyond $M$ will only lengthen $T_N$ without contributing to $Y_N$, leading to a decrease in system throughput $\frac{\mathrm{E}_\mathbf{X}[Y_N]}{\mathrm{E}_\mathbf{X}[T_N]}$. Hence, $\theta$ should always be set to a value that is \emph{no larger than} $M$ if $\frac{\mathrm{E}_\mathbf{X}[Y_N]}{\mathrm{E}_\mathbf{X}[T_N]}$ is to be maximized.
\end{remark}

\section{Throughput Performance}

\subsection{Distributions of $N^*$ and $\sum_{i=1}^{N^*}X_i$}
We analyze the distributions of $N^*$ and $\sum_{i=1}^{N^*}X_i$ in this section.
\begin{lemma}\label{lem:sumX-distribution}
For a given $n$ and an attempt rate $\lambda$, $\sum_{i=1}^{n}X_i$ is distributed as follows
\begin{eqnarray}\label{eqn:sumX-distribution}
&&\Pr\{\sum_{i=1}^{n}X_i=s\}\\
&=&\begin{cases}
\frac{1}{(e^\lambda-1)^n}\big(\sum_{j=M+1}^\infty\frac{\lambda^j}{j!}\big)^n & s=0\\
\frac{\lambda^s}{s!(e^\lambda-1)^n}\sum_{l=1}^n\binom{n}{l}\big(\sum_{j=M+1}^\infty\frac{\lambda^j}{j!}-1\big)^{n-l}l^s & s>0
\end{cases}.\nonumber
\end{eqnarray}
\end{lemma}
\noindent\emph{Proof}: See Appendix \ref{app:1}.

\begin{lemma}\label{lem:number-rounds}
Given a threshold $\theta\in (0,M]$ and an attempt rate $\lambda$, $N^*$ obtained by the optimal stopping rule \eqref{eqn:optimalstopping} has the distribution given in \eqref{eqn:rounds-distribution} at the top of the next page.

\begin{table*}[t!]
\begin{eqnarray}\label{eqn:rounds-distribution}
&&\Pr\{N^*(\lambda,\theta)=n\} =\\
&&\begin{cases}
\sum_{k=\theta}^M\frac{\lambda^k}{k!(e^\lambda-1)} & n=1\\
\frac{\sum_{i=\theta}^M\frac{\lambda^i}{i!}}{(e^\lambda-1)^n}\big(\sum_{i=M+1}^\infty\frac{\lambda^i}{i!}\big)^{n-1}+\frac{1}{(e^\lambda-1)^n}\sum_{s=1}^{\theta-1}\big(\sum_{i=\theta-s}^M\frac{\lambda^i}{i!}\big)\frac{\lambda^s}{s!}\sum_{l=1}^{n-1}\binom{n-1}{l}\big(\sum_{j=M+1}^\infty\frac{\lambda^j}{j!}-1\big)^{n-1-l}l^s & n>1\nonumber
 \end{cases}
\end{eqnarray}
\end{table*}
\end{lemma}
\noindent\emph{Proof}:
When $n=1$,
\begin{equation}
\Pr\{N^*(\lambda,\theta)=1\}=\Pr\{X_1\geq\theta\}=\sum_{k=\theta}^M\frac{\lambda^k}{k!(e^\lambda-1)},
\end{equation}
which proves the first half of \eqref{eqn:rounds-distribution}.
When $n>1$,
\begin{eqnarray}\label{eqn:n>1}
&&\Pr\{N^*(\lambda,\theta)=n\}=\Pr\big\{\sum_{i=1}^{n-1}X_i<\theta, \sum_{i=1}^{n}X_i\geq\theta\big\}\nonumber\\
&=&\sum_{s=0}^{\theta-1}\Pr\big\{X_n\geq \theta-s \big| \sum_{i=1}^{n-1}X_i=s\big\}\Pr\big\{\sum_{i=1}^{n-1}X_i=s\big\}\nonumber\\
&=&\sum_{s=0}^{\theta-1}\Pr\big\{X\geq \theta-s \big\}\Pr\big\{\sum_{i=1}^{n-1}X_i=s\big\},
\end{eqnarray}
where the last equality is due to the fact that the sequence of $X_i$ is i.i.d. Substituting \eqref{eqn:sumX-distribution} into \eqref{eqn:n>1}, the second half of \eqref{eqn:rounds-distribution} is obtained.$\hfill\blacksquare$

From Lemma \ref{lem:number-rounds}, we can  derive $\mathrm{E}[N^*(\lambda,\theta)]$ as
\begin{eqnarray}\label{meanN}
&&\mathrm{E}[N^*(\lambda,\theta)]=\sum_{n=1}^\infty n \Pr\{N^*(\lambda,\theta)=n\}\\
&=&\bigg(\sum_{k=\theta}^M\frac{\lambda^k}{k!(e^\lambda-1)}\bigg)\frac{1}{(1-\sum_{k=M+1}^\infty\frac{\lambda^k}{k!(e^\lambda-1)})^2}\nonumber\\
&+&\sum_{n=1}^\infty\frac{n+1}{(e^\lambda-1)^{n+1}}\sum_{l=1}^n\binom{n}{l}\bigg(\sum_{j=M+1}^\infty\frac{\lambda^j}{j!}-1\bigg)^{n-l}\nonumber\\
&&\times\sum_{k=1}^M\frac{\lambda^k}{k!}\sum_{s=\max(\theta-k,1)}^{\theta-1}\frac{\lambda^sl^s}{s!}.\nonumber
\end{eqnarray}

\begin{lemma}\label{lem:sumX-stopped}
 $\sum_{i=1}^{N^*(\lambda,\theta)}X_i$ follows the distribution given in \eqref{eqn:sumX-stopped}, shown at the top of the next page,  when the contention phase is stopped according to the stopping rule \eqref{eqn:optimalstopping}.
\begin{table*}[t!]
\begin{eqnarray}\label{eqn:sumX-stopped}
\Pr\big\{\sum_{i=1}^{N^*(\lambda,\theta)}X_i=s\big\}=
\begin{cases}
0 & s<\theta\\
\frac{\lambda^s}{s!(e^\lambda-1)}\bigg(\frac{1}{\sum_{j=1}^M\frac{\lambda^j}{j!(e^\lambda-1)}} & s\geq\theta\\
+\sum_{n=1}^\infty\frac{1}{(e^\lambda-1)^n}\sum_{l=1}^n\binom{n}{l}\big(\sum_{j=M+1}^\infty\frac{\lambda^j}{j!}-1\big)^{n-l}\sum_{t=1}^{\theta-1}\binom{s}{t}l^t\bigg)
\end{cases}
\end{eqnarray}
\end{table*}
\end{lemma}
\noindent\emph{Proof}:
For $s\geq \theta$, we have
\begin{eqnarray}\label{eqn:ProofsumX-stopped}
&&\Pr\big\{\sum_{i=1}^{N^*(\lambda,\theta)}X_i=s\big\}\\
&=&\sum_{n=1}^\infty\Pr\big\{\sum_{i=1}^nX_i=s\big|\sum_{i=1}^{n-1}X_i<\theta\big\}\Pr\big\{\sum_{i=1}^{n-1}X_i<\theta\big\}\nonumber\\
&=&\sum_{n=1}^\infty\sum_{t=0}^{\theta-1}\Pr\big\{X=s-t\big\}\Pr\big\{\sum_{i=1}^{n-1}X_i=t\big\}\nonumber\\
&=&\Pr\big\{X=s\big\}+\sum_{n=2}^\infty\sum_{t=0}^{\theta-1}\Pr\big\{X=s-t\big\}\Pr\big\{\sum_{i=1}^{n-1}X_i=t\big\}\nonumber
\end{eqnarray}
Substituting \eqref{eqn:X-distribution} and \eqref{eqn:sumX-distribution} into \eqref{eqn:ProofsumX-stopped}, we get \eqref{eqn:sumX-stopped}.$\hfill\blacksquare$

\subsection{Throughput of MPR WLAN with Multi-round Contention}
Given $\lambda$ and $\theta$, system throughput is calculated as
\begin{eqnarray}\label{eqn:throughput}
S(\lambda,\theta)&=&\frac{\mathrm{E}_\mathbf{X}\big[Y_{N^*(\lambda,\theta)}\big]}{\mathrm{E}_\mathbf{X}\big[T_{N^*(\lambda,\theta)}\big]}\\
&=&\frac{\mathrm{E}_\mathbf{X}\big[\min(\sum_{i=1}^{N^*(\lambda,\theta)} X_i,M)\big]}{\mathrm{E}_{\mathbf{X}}\big[N^*(\lambda,\theta)\big](T_{RTS}+m_I\sigma)+B} \text{ packets/sec}. \nonumber
\end{eqnarray}
where $\mathrm{E}_{\mathbf{X}}\big[N^*(\lambda,\theta)\big]$ is given by \eqref{meanN} and $\mathrm{E}_\mathbf{X}\big[\min(\sum_{i=1}^{N^*(\lambda,\theta)} X_i,M)\big]$ can be calculated as
\begin{eqnarray}\label{eqn:30}
\mathrm{E}_\mathbf{X}\big[\min(\sum_{i=1}^{N^*(\lambda,\theta)} X_i,M)\big]&=&\sum_{s=\theta}^M s\Pr\big\{\sum_{i=1}^{N^*(\lambda,\theta)}X_i=s\big\}\\
&+&\sum_{s=M+1}^\infty M\Pr\big\{\sum_{i=1}^{N^*(\lambda,\theta)}X_i=s\big\}\nonumber
\end{eqnarray}
WLAN throughputs studied in previous papers can be regarded as special cases of \eqref{eqn:throughput}. In particular, when $\theta=1$, \eqref{eqn:throughput} reduces to the throughput performance of MPR WLANs with single-round contention \cite{Zhang:09}. When $M=1$ and $\theta=1$, \eqref{eqn:throughput} reduces to the throughput of traditional WLANs with single-packet reception \cite{Bianchi:00}.

From \eqref{eqn:vn}, it can be seen that there exists an optimal $\theta^*(\lambda)$ that maximizes the system throughput for a given $\lambda$ (or equivalently, a given distribution of $X$). With the analysis in this section, $\theta^*(\lambda)$ can be obtained by performing a simple line search instead of calculating directly from \eqref{eqn:vn}. In addition, if we have the freedom to adjust the attempt rate $\lambda$ as well, the maximum system throughput can be achieved by jointly optimizing $\lambda$ and $\theta$:
\begin{equation}\label{eqn:optS}
S^*=\max_{\lambda,\theta}S(\lambda,\theta).
\end{equation}
%

If we blindly set $\theta=M$ without optimizing it, then we obtain a lower bound
\begin{eqnarray}
B_L(\lambda)&\triangleq& S(\lambda,M)=\frac{M}{\mathrm{E}_{\mathbf{X}}\big[N^*(\lambda,M)\big](T_{RTS}+m_I\sigma)+B}\nonumber\\
&\leq& S(\lambda,\theta^*(\lambda)),
\end{eqnarray}
which yields
\begin{eqnarray}\label{eqn:BL}
B_L^*=\max_\lambda B_L(\lambda)\leq S^*.
\end{eqnarray}

As we will show shortly, the gap between $B_L^*$ and $S^*$ is marginal in most cases. Nonetheless, $B_L(\lambda)$ is much easier to calculate than $S(\lambda,\theta)$. Moreover, to achieve $B_L^*$, we simply need to find the $\lambda$ that minimizes $\mathrm{E}_{\mathbf{X}}[N^*(\lambda,M)]$. This is much less computationally involved than finding the right $\lambda$ and $\theta$ to maximize $S(\lambda,\theta)$.  Hence, $B_L^*$ serves a good approximation of $S^*$ from both analytical and practical perspective.

\begin{figure}[h!]
\centering
\includegraphics[width=0.5\textwidth]{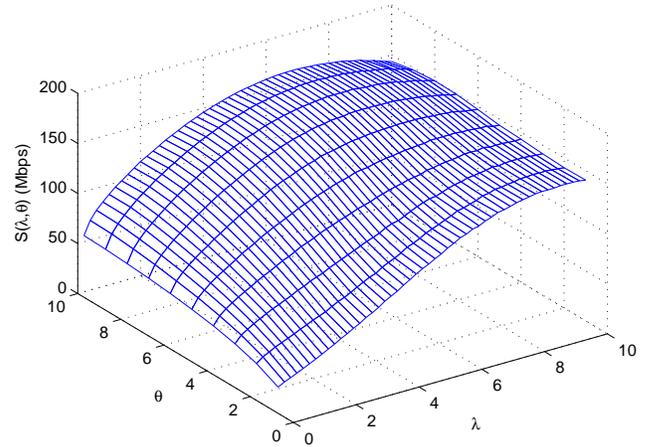}
\caption{$S(\lambda,\theta)$ for IEEE 802.11a with $M=10$, $L=8184$ bits and data transmission rate 54 Mbps.}\label{fig:3}
\end{figure}

 As an illustration, throughput $S$ (in unit of Mbps) is plotted against $\lambda$ and $\theta$ in Fig. \ref{fig:3} for an IEEE 802.11a WLAN with $M=10$ and data transmission rate 54 Mbps. Other system parameters are shown in Table \ref{tab:parameters}. It can be seen that for each attempt rate $\lambda$, there exists an optimal $\theta$ that maximizes throughput $S$, and vice versa. In general, traditional single-round contention (i.e., setting $\theta=1$) does not yield as high throughput as multi-round contention (i.e., setting $\theta>1$).

\begin{table}
\caption{System Parameters (Adopted from IEEE 802.11a)}
\centering
\begin{tabular}{|c|c|}
  \hline
  Packet payload & 8184 bits \\
  \hline
  MAC header & 224 bits \\
  \hline
  PHY overhead & 20 $\mu s$+22/6$\mu s$ \\
  \hline
  ACK & 112 bits + PHY overhead \\
  \hline
  RTS & 160 bits + PHY overhead \\
  \hline
  CTS & 112 bits + PHY overhead \\
  \hline
  Basic rate & 6 Mbps \\
  \hline
  Data rate & 54 Mbps \\
  \hline
  Slot time $\sigma$ & 9 $\mu s$ \\
  \hline
  SIFS & 16 $\mu s$ \\
  \hline
  DIFS & 34 $\mu s$ \\
  \hline
\end{tabular}\label{tab:parameters}
\end{table}

\begin{figure}[h!]
\centering
\includegraphics[width=0.5\textwidth]{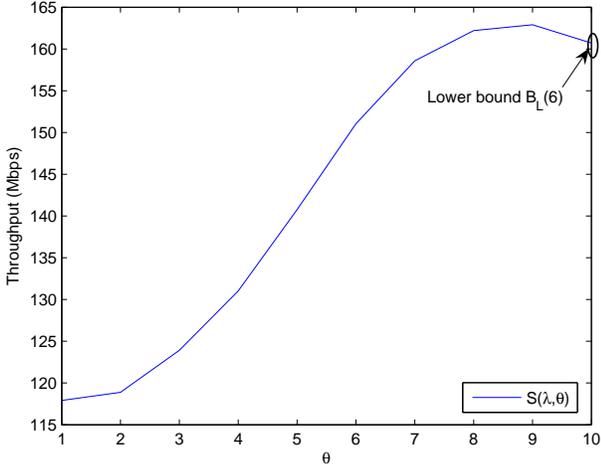}
\caption{$S(\lambda,\theta)$ and $B_L(\lambda)$ for IEEE 802.11a with $\lambda=6$ and $M=10$, $L=8184$ bits and data transmission rate 54 Mbps.}\label{fig:4}
\end{figure}

\begin{figure}[h!]
\centering
\includegraphics[width=0.5\textwidth]{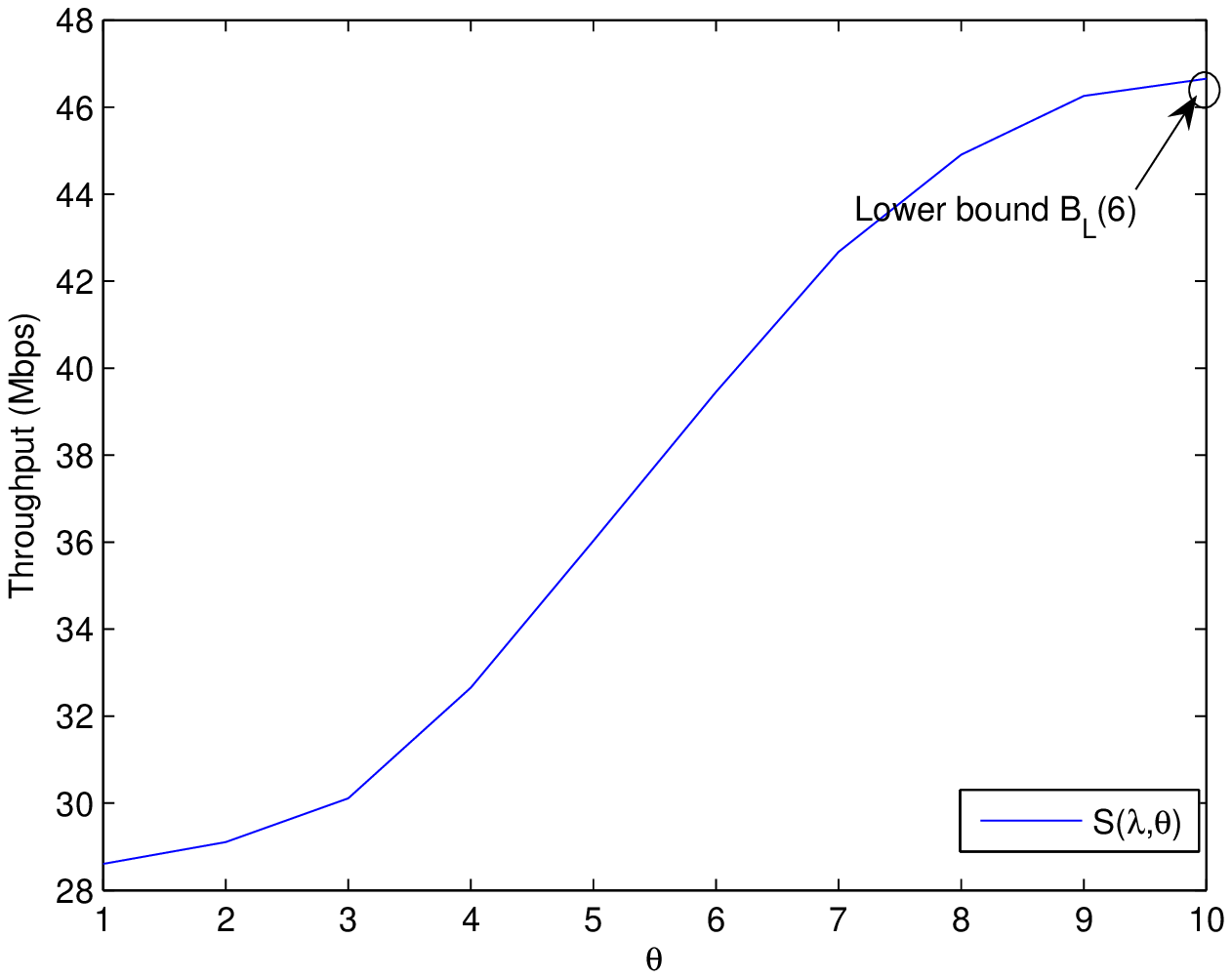}
\caption{$S(\lambda,\theta)$ and $B_L(\lambda)$ for IEEE 802.11a with $\lambda=6$ and $M=10$, $L=8184$ bits and data transmission rate 6 Mbps.}\label{fig:5}
\end{figure}

Fixing $\lambda=6$, we plot $S(\lambda,\theta)$ and $B_L(\lambda)$ in Fig. \ref{fig:4}. In particular, $B_L(\lambda)= S(\lambda,10)$ according to the definition. It can be seen that $B_L$ is close to the \emph{maximum} value of  $S(\lambda,\theta)$, which occurs when $\theta=9$. Furthermore, note that only $72\%$ of the maximum throughput can be achieved when $\theta=1$, in which case the system reduces to one with single-round contention. Reducing the data transmission rate to 6 Mbps, we plot the curve again in Fig. \ref{fig:5}. In this case, $B_L$ coincides with the maximum value of $S(\lambda,\theta)$.

\begin{figure}[h!]
\centering
\includegraphics[width=0.5\textwidth]{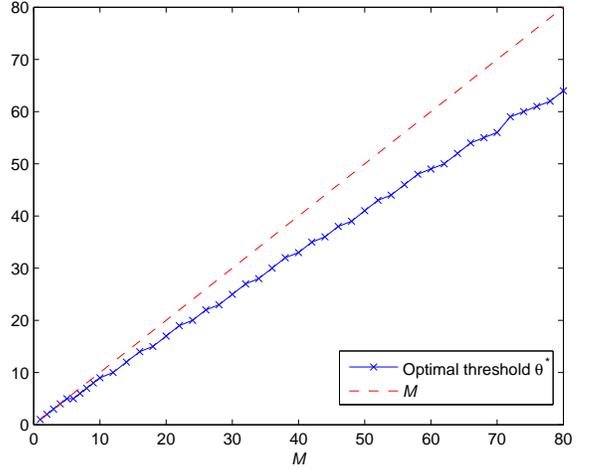}
\caption{Optimal $\theta^*$ for IEEE 802.11a with $L=8184$ bits and data transmission rate 54 Mbps.}\label{fig:8}
\end{figure}

In Fig. \ref{fig:8}, we plot the optimal threshold $\theta^*(\lambda^*)$ obtained from \eqref{eqn:optS}. One interesting observation is that $\theta^*$ is not necessarily close to $M$, especially for large $M$. This implies the optimal strategy would rather transmit fewer than $M$ packets than executing too many contention rounds in these cases.

\section{Throughput Scaling and Comparison with Single-Round Contention}
Our previous study in \cite{Zhang:09} has demonstrated MPR as a powerful capacity-enhancement technique in traditional WLANs with single-round contention. In particular, we have proved that the maximum throughput increases super-linearly with $M$, the MPR capability of the channel. In this section, we extend the study by investigating (i) how the maximum system throughput scales with the MPR capability $M$ under multi-round contention; and (ii) how multi-round contention improves system performance compared with single-round contention. In the following figures, $B_L^*$ is obtained through analytical approaches, while $S^*$ is obtained through semi-analytical simulations.

\begin{figure}[h!]
\centering
\includegraphics[width=0.5\textwidth]{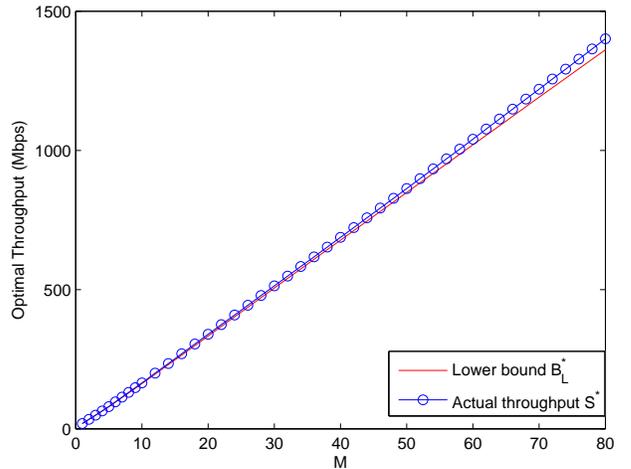}
\caption{$S^*$ and $B_L^*$ for IEEE 802.11a with $L=8184$ bits and data transmission rate 54 Mbps.}\label{fig:6}
\end{figure}

In Fig. \ref{fig:6}, we plot the maximum throughput $S^*$ and its lower bound $B_L^*$ as a function of $M$ when system parameters are set as in Table \ref{tab:parameters}. It can be seen that system throughput increases drastically with the increase of MPR capability. Moreover, the lower bound $B_L^*$ is very close to the actual throughput $S^*$.

\begin{figure}[h!]
\centering
\includegraphics[width=0.5\textwidth]{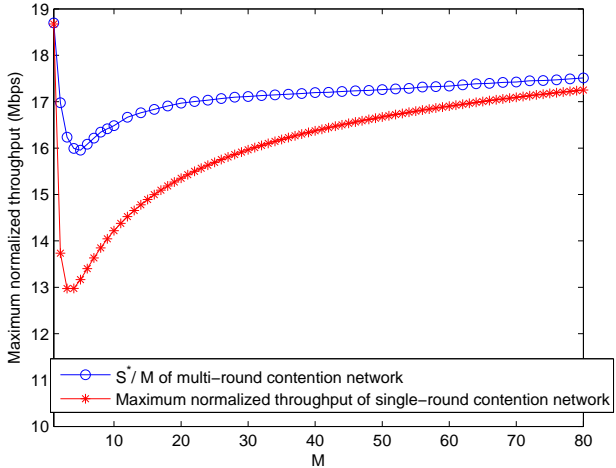}
\caption{Maximum normalized throughputs for IEEE 802.11a with $L=8184$ bits and data transmission rate 54 Mbps.}\label{fig:7}
\end{figure}

The maximum normalized throughput with respect to $M$, i.e., $\frac{S^*}{M}$, is plotted in Fig. \ref{fig:7}. For comparison, the maximum normalized throughput of single-round contention RTS/CTS access network (derived in \cite{Zhang:09}) is also plotted. Three conclusions can be drawn from the figure. First, similar to the single-round contention case, $\frac{S^*}{M}$ increases with $M$ in the multi-round contention case when $M$ is larger than 4. In other words, multi-round contention preserves the super-linear throughput scaling. In practical systems, $M$ is directly related to the cost (e.g., bandwidth in CDMA systems or the number of antennas in multi-antenna systems). Super-linear scaling of throughput implies that the achievable throughput per unit cost increases with $M$. Second, multi-round contention significantly improves system throughput compared with single-round contention, especially for small to medium $M$ (say $M\leq 20$). The throughput improvement can be as high as $23\%$. This is because the channel is more likely to be ``fully occupied" with packets with the multi-round contention MAC. Note that small to medium $M$ is of particular interest for practical applications, where the multiuser detection capability at the receiver is typically not high. This provides a strong incentive for the deployment of a multi-round contention MAC in future wireless networks. Third, the gap between the normalized throughputs of multi-round contention and single-round contention networks diminishes when $M$ grows (perhaps impractically) large. This is not surprising, however, as we have proved in \cite{Zhang:09} that the throughput penalty due to distributed random access diminishes to zero when $M$ becomes large even with single-round contention. Thus, the optimal stopping strategy we have derived may turn out to stop the contention process after one contention round most of the time.

Before leaving this section, note that we have assumed that the attempt rate $\lambda$ remains constant for all contention rounds. In principle, throughput $S^*$ can be further improved by allowing $\lambda$ to vary from one contention round to another. For example, a smaller $\lambda$ should be adopted if the number of winning stations is already close to the threshold $\theta$ so as to reduce the probability of collision in the contention round.  By doing so, however, the derivation of the optimal stopping rule would be much more involved. Fortunately, from the throughput upper bound that we  derive in Appendix \ref{app:2}, it can be seen that the potential throughput enhancement by varying $\lambda$ across different slots is marginal.

\section{Discussions and Verification of Analysis}
In this section, we verify the analysis through simulation. We also discuss the validity of the Poisson assumption adopted in Sections IV and V.

\begin{figure}[h!]
\centering
\includegraphics[width=0.5\textwidth]{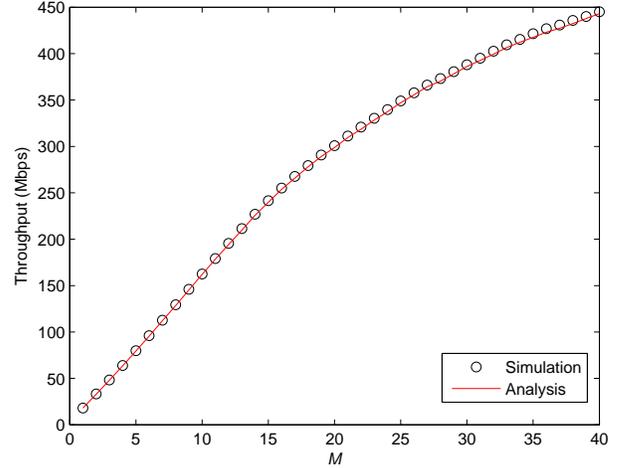}
\caption{Simulation. $r=2$, $W_0=16$, $K=100$.} \label{fig:100station}
\end{figure}

In Fig. \ref{fig:100station}, we simulate the multi-round IEEE 802.11 WLAN described in Section II-B and Fig. \ref{fig:802.11multiround} when there are $K=100$ stations. In the figure, the MPR capability $M$ varies from 1 to 40. We set backoff exponent $r=2$, minimum contention window $W_0=16$, and threshold $\theta=M$. Other parameters are the same as in Table \ref{tab:parameters}. For each $M$, the simulation is run for $100,000$ generic time slots after $5,000$ slots of warm-up. For comparison, we also plot the analytical results $S(\lambda,\theta)$ by setting $\lambda$ to be the average aggregate transmission probability obtained from the simulations\footnote{For given $K$ and $M$, the transmission probability is determined by backoff parameters such as $r$ and $W_0$.}. It can be seen from the figure that the simulation and analytical results almost overlap when $M$ is relatively small. When $M$ exceeds 30, the simulation result deviates slightly from the analysis. This is because for large $M$, each station tends to transmit at a higher probability $\tau$ under the exponential backoff scheme. In this case, $\lambda=K\tau$ is not much smaller than $K$, and hence the Poisson assumption becomes less accurate.

\section{Conclusions}
In this paper, we have proposed a multi-round contention random-access protocol for WLANs with MPR capability. An optimal stopping rule is derived to strike the desired tradeoff between channel utilization and contention overhead. In particular, we prove that the one-stage look-ahead rule, which is a simple threshold-based rule, is optimal due to the special feature of the return function. The multi-round contention protocol significantly improves system throughput compared with conventional single-round contention protocols, especially for small to medium $M$. This is because the MPR channel is now more likely to be packed with as many packets as it can resolve. Furthermore, multi-round contention preserves super-linear throughput scaling, providing a strong incentive to deploy MPR in future WLANs.

In Sections IV and V, we have assumed that $K$ is large enough so that the number of transmissions in a generic time slot follows a Poisson distribution. Our simulation in Fig. \ref{fig:100station} shows that this assumption is very accurate when $K$ is sufficiently larger than $M$. On the other hand, the Poisson assumption is less accurate when $K$ is relatively small. In this case, the transmission attempts follows a binomial distribution that varies from one contention round to another, as the number of potential contenders decreases with the number of contention rounds. As a result, $\theta_n$ defined in \eqref{eqn:1slathreshold} is no longer invariant with $n$, making the analysis of the 1-sla scheme much more complicated. In our future work, we will devise effective mechanisms to analyze multi-round contention MPR WLANs for small to medium $K$.

\appendices
\section{Proof of Lemma \ref{lem:sumX-distribution}}\label{app:1}
The proof is trivial for $s=0$.

 For $s>0$, we prove the lemma by induction. It is obvious that Lemma \ref{lem:sumX-distribution} holds when $n=1$. Assuming that Lemma \ref{lem:sumX-distribution} holds when $n=N$, we show in \eqref{eqn:proofsum} that it also holds for $n=N+1$ in the following.
\begin{table*}[t!]
\begin{eqnarray}\label{eqn:proofsum}
&&\Pr\{\sum_{i=1}^{N+1}X_i=s\}=\sum_{s_N=1}^{s-1}\Pr\{X_{N+1}=s-s_N\big|\sum_{i=1}^{N}X_i=s_N\}\Pr\{\sum_{i=1}^{N}X_i=s_N\}\nonumber\\
&&~~+\Pr\{X_{N+1}=0\big|\sum_{i=1}^{N}X_i=s\}\Pr\{\sum_{i=1}^{N}X_i=s\}+\Pr\{X_{N+1}=s\big|\sum_{i=1}^{N}X_i=0\}\Pr\{\sum_{i=1}^{N}X_i=0\}\nonumber\\
&=&\sum_{s_N=1}^{s-1}\frac{\lambda^{s-s_N}}{(s-s_N)!(e^\lambda-1)}\frac{\lambda^{s_N}}{s_N!(e^\lambda-1)^N}\sum_{l=1}^N\binom{N}{l}\big(\sum_{j=M+1}^\infty\frac{\lambda^j}{j!}-1\big)^{N-l}l^{s_N}\nonumber\\
&&~+\sum_{j=M+1}^\infty\frac{\lambda^j}{j!(e^\lambda-1)}\frac{\lambda^s}{s!(e^\lambda-1)^N}\sum_{l=1}^N\binom{N}{l}\big(\sum_{j=M+1}^\infty\frac{\lambda^j}{j!}-1\big)^{N-l}l^s+\frac{\lambda^s}{s!(e^\lambda-1)}\frac{1}{(e^\lambda-1)^N}\big(\sum_{j=M+1}^\infty\frac{\lambda^j}{j!}\big)^N\nonumber\\
&=&\frac{\lambda^s}{s!(e^\lambda-1)^{N+1}}\bigg\{\sum_{l=1}^N\binom{N}{l}\big(\sum_{j=M+1}^\infty\frac{\lambda^j}{j!}-1\big)^{N-l}((l+1)^s-l^s-1)\nonumber\\
&&~+\big(\sum_{j=M+1}^\infty\frac{\lambda^j}{j!}\big)\sum_{l=1}^N\binom{N}{l}\big(\sum_{j=M+1}^\infty\frac{\lambda^j}{j!}-1\big)^{N-l}l^s+\big(\sum_{j=M+1}^\infty\frac{\lambda^j}{j!}\big)^N\bigg\}\nonumber\\
&=&\frac{\lambda^s}{s!(e^\lambda-1)^{N+1}}\sum_{l=1}^{N+1}\binom{N+1}{l}\big(\sum_{j=M+1}^\infty\frac{\lambda^j}{j!}-1\big)^{N+1-l}l^s,
\end{eqnarray}
\end{table*}

\section{Multi-Round Contention with Carry-Over: A Throughput Upper Bound}\label{app:2}
In our proposed scheme, a winning station may not be selected for data transmission when there are more than $M$ winning stations by the end of the contention phase. The unselected stations regard themselves as having encountered virtual collisions and back off.

In this appendix, we propose an alternative scheme where unselected winning stations are carried over to the next super round instead of being discarded. In other words, these stations are automatically categorized as winning stations without the need to contend for the channel again. This scheme is based on an ideal assumption that the AP can memorize the contention outcomes of the previous round. Hence, no ``contention efforts" are wasted. All winning stations can eventually transmit without the need to contend again. As such, it is always optimal to wait until there are no fewer than $M$ winning stations (including the carried-over ones) before data transmission. The system throughput is given by
\begin{eqnarray}\label{eqn:Sc}
S_c=\frac{M}{\frac{M}{\mathrm{E}[X]}(T_{RTS}+m_I\sigma)+B} \text{ packets/second}.
\end{eqnarray}
with the optimal $\lambda$ that maximizes $S_c$ being
\begin{eqnarray}
\lambda_c^*=\arg\max S_c=\arg\max \frac{\mathrm{E}[X]}{T_{RTS}+m_I\sigma}.
\end{eqnarray}
It is not surprising that the optimal $\lambda_c^*$ is simply the one that maximizes the number of winning stations per unit time during the contention phase.

\begin{figure}[h!]
\centering
\includegraphics[width=0.5\textwidth]{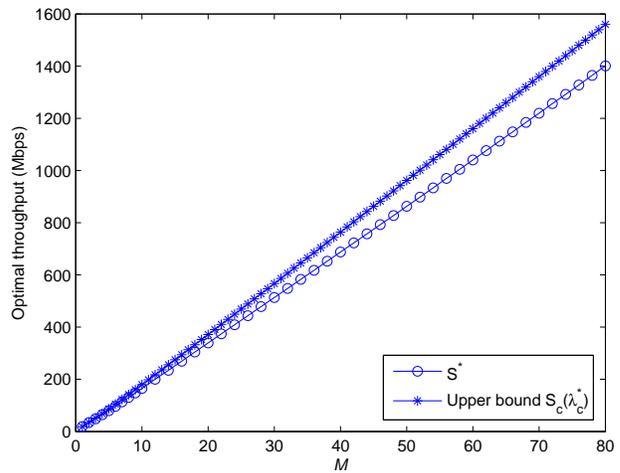}
\caption{Throughput upper bound for IEEE 802.11a with $L=8184$ bits and data transmission rate 54 Mbps.}\label{fig:9}
\end{figure}

$S_c(\lambda_c^*)$ serves as an upper bound of the throughput of multi-round contention WLAN without carry over. In other words, it puts a cap on the potential throughput enhancement by varying $\lambda$ across contention rounds. In Fig. \ref{fig:9}, the throughput upper bound is plotted together with the maximum throughput $S^*$ of the non-carry-over protocol. It shows that $S^*$ can hardly be further improved when $M$ is small, and at most by $11\%$ when $M$ is as large as 80.

\begin{biography}[{\includegraphics[width=1in,height=1.25in,clip,keepaspectratio]{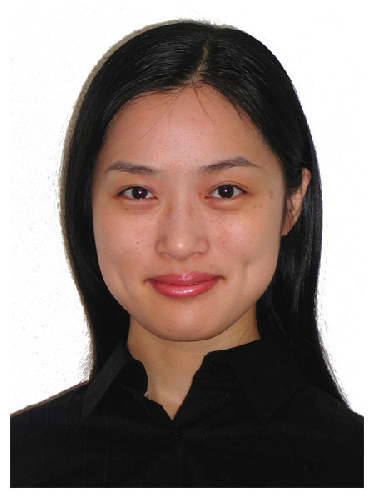}}]{Ying Jun (Angela) Zhang}
(S'00, M'05) received here PhD degree in Electrical and Electronic Engineering from the Hong Kong University of Science and Technology, Hong Kong in 2004. Since Jan. 2005, she has been with the Department of Information Engineering in The Chinese University of Hong Kong, where she is currently an assistant professor.

Dr. Zhang is on the Editorial Boards of IEEE Transactions of Wireless Communications and Willey Security and Communications Networks Journal. She has served as a TPC Co-Chair of Communication Theory Symposium of IEEE ICC 2009, Track Chair of ICCCN 2007, and Publicity Chair of IEEE MASS 2007. She has been serving as a Technical Program Committee Member for leading conferences including IEEE ICC, IEEE GLOBECOM, IEEE WCNC, IEEE ICCCAS, IWCMC, IEEE CCNC, IEEE ITW, IEEE MASS, MSN, ChinaCom, etc. Dr. Zhang is an IEEE Technical Activity Board GOLD Representative, 2008 IEEE GOLD Technical Conference Program Leader, IEEE Communication Society GOLD Coordinator, and a Member of IEEE Communication Society Member Relations Council (MRC).

Her research interests include wireless communications and mobile networks, adaptive resource allocation, optimization in wireless networks, wireless LAN/MAN, broadband OFDM and multicarrier techniques, MIMO signal processing.
As the only winner from Engineering Science, Dr. Zhang has won the Hong Kong Young Scientist Award 2006, conferred by the Hong Kong Institution of Science.

\end{biography}

\end{document}